\documentclass[preprint,12pt]{revtex4}

\usepackage{amssymb,amsfonts,amsmath}
\usepackage{graphicx}

\newcommand{\correction}[1]{#1}

\begin{document}


\title{Quantifying reflexivity in financial markets: towards a prediction of flash crashes}

\author{Vladimir Filimonov}
\email[]{vfilimonov@ethz.ch}
\affiliation{Department of Management, Technology and Economics ETH Z\"{u}rich}

\author{Didier Sornette}
\email[]{dsornette@ethz.ch}
\affiliation{Department of Management, Technology and Economics ETH Z\"{u}rich
and Swiss Finance Institute, c/o University of Geneva}

\date{\today}

\begin{abstract} 
We introduce a new measure of activity of financial markets that 
provides a direct access to their level of endogeneity. This measure quantifies how much of price changes
are due to endogenous feedback processes, as opposed to exogenous news. 
For this, we calibrate the self-excited conditional Poisson Hawkes model, which combines in a natural and parsimonious way exogenous influences with self-excited dynamics, to the E-mini S\&P 500 futures contracts traded in the Chicago Mercantile Exchange from 1998 to 2010. We find that the level of endogeneity has increased significantly from 1998 to 2010, with only 70\% in 1998 to less than 30\% since 2007 of the price changes resulting from some revealed exogenous information. Analogous to nuclear plant safety concerned with avoiding ``criticality'', 
our measure provides a direct quantification of the distance of the financial market to
a critical state defined precisely as the limit of diverging
trading activity in absence of any external driving. 

\end{abstract}



\maketitle

\section{Introduction}

Human societies and more generally biological communities are characterized by 
significant levels of interactions developing into complex network of interdependencies,
leading to the emergence of remarkable dynamical properties and abilities. 
These societies are however not just evolving spontaneously or endogenously.
In absence of external ``forces'', they would freeze, regress or die.
Essential to their organization is the flux of exogenous influences 
that allow them to ``feed'', adapt, learn and evolve. 
For instance, both external stimuli and endogenous collective and interactive wiring
between neurons are essential for brain development and continuous performance. 
Scientific discoveries result from the interplay between the maturation of ideas
and technical progress within the scientific community
as well as serendipity \cite{RobertMerton1961,Buchanan2009}.
One can find many other examples of the essential interplay between
endogenous processes and exogenous dynamics in interesting out-of-equilibrium
systems, that is, almost everywhere from the micro to the macro worlds~\cite{Sorendoexorev}.

A fundamental question is: how much of the observed dynamics is due to the 
external influences versus internal processes? Is it possible to quantify the interplay
between exogeneity and endogeneity? Can this be used for characterizing the robustness
of systems and for developing diagnostics of fragility and of incoming crises
as well as upside potentials?

Here, to address these questions quantitatively,
we consider financial markets as paradigms of complex human societies,
in which external news play the role of the exogenous influences impacting 
investors whose elaborate interactions via complex social and economic networks
lead to price formation. In a nutshell, one can indeed state that 
financial markets are nothing but the engines through which information is transformed into prices.
But, how much of the news and information are really captured by prices? 
Is it not the case that prices also reflect idiosyncratic dynamics of social networks that may
lead to deviations from true valuation?
The arguably most important question in financial economics is indeed to what degree
are prices faithful embodiments of information? And, correlatively,  what is ``information''?


Financial economists have introduced the ``Efficient Market Hypothesis'' (EMH) that \correction{states, in its 
ideal limit, that the market absorbs in full and essentially instantaneously 
the flow of information by faithfully reflecting it in asset prices}~\cite{Fama1,Fama2,Samuelson1,Samuelson2}.  
The EMH amounts to considering the \correction{process of price formation} as 
perfect with almost instantaneous reaction and infinite precision, so that only 
external influences show up. In other words, according to the EMH, prices are just reflecting news: 
\correction{whatever the internal structure of the market, the EMH assumes that the market
is sufficiently fast and effective so that its converges to an equilibrium price justly 
reflecting exogenous information, while all endogenous processes have had time
to converge after each given exogenous shock, thus disappearing from the observations.
As a result, markets are assumed to be driven only by an external inputs of information
and only reflect them. Only updates of these inputs can change investors' anticipation and thus prices.
In particular, such extreme events as financial crashes are, according to the EMH, the signature 
of exogenous negative news of large impact.}

In reality, it is now recognized that prices move much too much 
compared with what would be expected from the EMH (even corrected
for the costs of gathering information), i.e., from
the volatility of fundamental news proxied for instance by dividends
\cite{Shiller1981,LeRoy1981}. Moreover, 
according to the EMH, large price moves should only occur with
significant geo-politico-economic-financial news. This prediction has been refuted by diverse studies comparing 
price movements and relevant news at the daily time scale \cite{Cutler1987}
and for high frequency financial data \cite{Bouchaud2008,Bouchaud2010_endo}, 
which showed that only a small fraction of price movements could be explained by relevant news releases. 
This suggests that price dynamics are mostly endogenous and driven by positive feedback mechanisms
involving investors' anticipations that lead to self-fulfilling prophecies, as described
qualitatively by Soros' concept of ``market reflexivity''~\cite{Soros_Alchemy1988}.

The contribution of the present work is to provide what is, to the best of our knowledge, the first quantitative
estimate of the degree of reflexivity, measured as the proportion of price moves due to 
endogenous interactions to the total number of all price moves that also include the impact of exogenous news.
We use the self-excited conditional Poisson
Hawkes model \cite{Hawkes1971}, which combines in a natural and parsimonious way exogenous influences with self-excited dynamics.
According to the Hawkes model, each event, i.e. price change, may lead to a whole tree of offsprings, i.e. other price changes. 
The Hawkes model provides a natural set-up to describe the endogenous mechanisms resulting from herding 
as well as from strategic order splitting, which lead to long-range correlation in the series of trade initiations (buy-side or sell-side). 
The self-excited Hawkes branching process allows us to classify different types of volatility shocks and 
to separate the exogenous shocks from the endogenous dynamics. 
In particular, 
the Hawkes model provides a single parameter, the so-called ``branching ratio'' `$n$' that measures directly
the level of endogeneity. The parameter $n$ can be interpreted as the 
fraction of endogenous events within the whole population of price changes. Our observations of this measure of endogeneity reflect a robust behavioral trait of human beings who tend to herd more 
at short time scales in time of fear and panic. Our study thus complements
the evidence for herding at the time scales of years over with financial bubbles
develop~\cite{Sornettecrashbook}, by showing the existence of herding at short time scales
according to a different mechanism than the ones operating at large time scales.

\section{The Data}

We use the E-mini S\&P 500 futures contracts (ticker-symbol ES), which are traded in the Chicago Mercentile Exchange. Being introduced in 1997 as a supplement to the regular S\&P 500 futures contracts with reduced contract sizes, E-mini's had attracted a lot of small investors and has became one of the most actively traded contracts in the world. Our dataset contains all transactions and changes in supply and demand (best bid and best ask sides) from January 5, 1998 to August 29, 2010 (in total 2'431'967'666 records including 298'586'423 transactions), which are recorded with corresponding volume and timestamps that are rounded to the nearest second. The dataset was cleaned of gaps and non-trading days. At every moment, there are traded futures contracts with 5 different maturities \correction{in the March Quarterly Cycle (with expirations on the third Friday of March, June, September and December). However the rollover dates are 8 days before the expiration dates, i.e. on the second Thursdays of each of these months. At the rollover date, the liquidity (measured in volume) of the contract that is going to expire is switched to the contract that will expire at the following quarter. For the analyzed dataset, the volume of the most actively traded contracts accounts for 96.7\% of the total traded volume (99.1\% when excluding rollover weeks).} Thus, in our studies, we have focused only on the most actively traded contracts --- the contract whose rollover date is closest to the given moment.

The analyzed dataset presents highly non stationary properties both at fine-grained and coarse-grained scales. At the daily scale, the trading activity has increased dramatically over these 10 years: starting  in 1998 with
an average of 7,000 transactions per day on 16,000 contracts, the average daily volume in 2009 consisted
in 148,000 transactions on approximately 1,837,000 contracts  (see fig.~\ref{fig_2}A). At the same time, during any given day, the activity is very low outside Regular Trading Hours (9:30--16:15 EST), and exhibits the well-known the U-shape intraday seasonality\correction{, which is illustrated in fig.~\ref{fig_4}A.}

At any given moment $t$, one may distinguish four different prices in the market, that reflect its different properties: (i) \emph{the last transaction price} $p_{tr}(t)$, at which the previous transaction was executed, (ii) \emph{the best ask price} $a(t)$ and (iii) \emph{the best bid price} $b(t)$ at which market participants may immediately correspondingly by and sell an asset, and (iv) \emph{the mid-price}, which is defined as the average of best bid and ask prices: $p_m(t)=(a(t)+b(t))/2$. The bid and ask prices reflect correspondingly demand and supply of the liquidity providers; \correction{the transaction price reflects actions of liquidity takers, and mid-price changes result from actions of all market participants: both liquidity providers and takers. The transactions are triggered when a market order arrives. In case of a buy market order, the transaction is executed at the best ask price, while a sell market order triggers a transaction at the best bid price. Since the sequence of order arrivals is stochastic with the sign of order being a random variable, the last transaction price will jump from best bid to best ask price and back even without changes in the balance between supply and demand. This stochastic behavior, which is called ``bid-ask bounce'', represents a kind of ``noise source'' to the price.}

\correction{The idea that the last transaction price in high frequency financial data is a poor proxy of the unobservable asset's value, which is subjected to the additive ``microstructure noise'', is a well established concept in the market microstructure literature (see for instance,~\cite{AitSahalia2005}, and the concept of ``noise traders'' by F. Black~\cite{Black1986}). In contrast to the last transaction price, the mid-price is free from the bid-ask-bounce and is changed only when the balance between supply (liquidity providers) and demand (liquidity takers) is upset. Therefore, the mid-price is claimed to be a better proxy for the asset value, given the information available~\cite{Hasbrouck1991,Engle2000}. 
In the ``market impact'' (or ``price impact'') literature that study the question of how much the price of an asset will change due to a single market order execution, the mid-price became the ``default measure'' of the price movements (see, for instance, extensive review~\cite{BouchaudLillo2009}).}

\correction{In the present study, we consider the changes of the mid-price of E-Mini S\&P 500 futures as the best proxy for market movements as a whole.} More precisely, we apply and test the Hawkes model to events corresponding to changes in mid-price of E-mini's within the Regular Trading Hours (in total 24'309'652 events).

\section{Self-excited Hawkes Model}

A class of models based on continuous or discrete stochastic price processes have built on Bachelier's random walk model~\cite{Bachelier1900}
to incorporate stylized facts, such as quasi-absence of linear autocorrelation of returns, the existence of long-memory in the volatility, volatility clustering and multifractality, fat tails in the distributions of returns, correlation between volatility and volume, time-reversal asymmetry, the leverage effect, gain-loss asymmetries and others (see for instance~\cite{Bouchaud_Risks2000,Cont2001}).

Another class of models view the price formation process as discontinuous, i.e., following point processes
reflecting the discrete nature of market order arrivals. The Poisson point process is the simplest of this class, 
in which events occur independently of one another with a constant arrival rate $\lambda$. 
Having no correlation structure, the Poisson point process cannot describe 
the stylized facts of real order flows, such as (i) clustering of order arrivals, (ii) long memory in inter-trade intervals~\cite{Ivanov2004,JiangZhou2009}, (iii) slower-than-exponential decay of the distribution of inter-trade intervals~\cite{Ivanov2004,Eisler2006,Politi2008}, (iv) long memory of the signs of successive trades~ \cite{BouchaudLillo2009} and (v) multifractal scaling of inter-trade intervals~\correction{\cite{JiangZhou2009,Oswiecimka2005,Perello2008}}. A first attempt to account partially for these 
stylized facts characterizing high frequency transaction data uses a class of 
self-excited processes called Autoregressive Conditional Durations (ACD) model~\cite{Engle1997,Engle1998}, which describes the inter-event durations with a GARCH-type equation. A more consistent approach based on the generalization of the self-excited Hawkes model~\cite{Hawkes1971} was developed in the working paper~\cite{Bowsher2002} (published later with corrections~\cite{Bowsher2007}), where 
the author used a bivariate Hawkes process to model the arrival times of market buy and sell orders. This approach 
later became the ``gold standard'' for the use of self-excited models to describe high frequency order flows~\cite{Hewlett2006,Bauwens2009} and was later extended to account for the actions of liquidity providers in the construction process of the order book ~\cite{Large2007,Toke2011,Cont2011}. Recently, the multivariate Hawkes process was used to model last transaction price data and, in particular, the signature plot and the Epps effect~\cite{Muzy2010hawkes,Muzy2011}.

The Hawkes point process can be regarded as the generalization of the non-homogeneous Poisson process, whose intensity $\lambda(t)$ (defined such that $\lambda(t) dt$ is the expected value of the number of events in the time interval $[t,t+dt)$) not only depends on time $t$ but also on the history of the process according to
\begin{equation}\label{hawkes}
	\lambda_t(t)=\mu(t)+\sum_{t_i<t}h(t-t_i)~,
\end{equation}
where $t_i$ are the timestamps of the events of the process, $\mu(t)$ is a \emph{background intensity} that accounts for \emph{exogenous} events (not dependent on history) and $h(t)$ is a \emph{memory kernel function} that weights how much
past events influence the generation of future events and thus controls the amplitude of the
\emph{endogenous} feedback mechanism. For our purposes, the Hawkes process presents two 
interesting properties. First, the external influences on the system ($\mu(t)$) and the internal feedback mechanisms 
($h(t)$) can be clearly isolated in their linear additive contributions to the conditional intensity $\lambda_t(t)$
in expression (\ref{hawkes}). Second, the linear structure $\lambda_t(t)$
in the Hawkes model allows one to map it exactly onto a \emph{branching process} \cite{Verejones}. This implies
that the process exhibits a critical bifurcation controlled by the so-called branching ratio parameter to be defined shortly below.

\begin{figure}[p]
\centerline{\includegraphics[width=\textwidth]{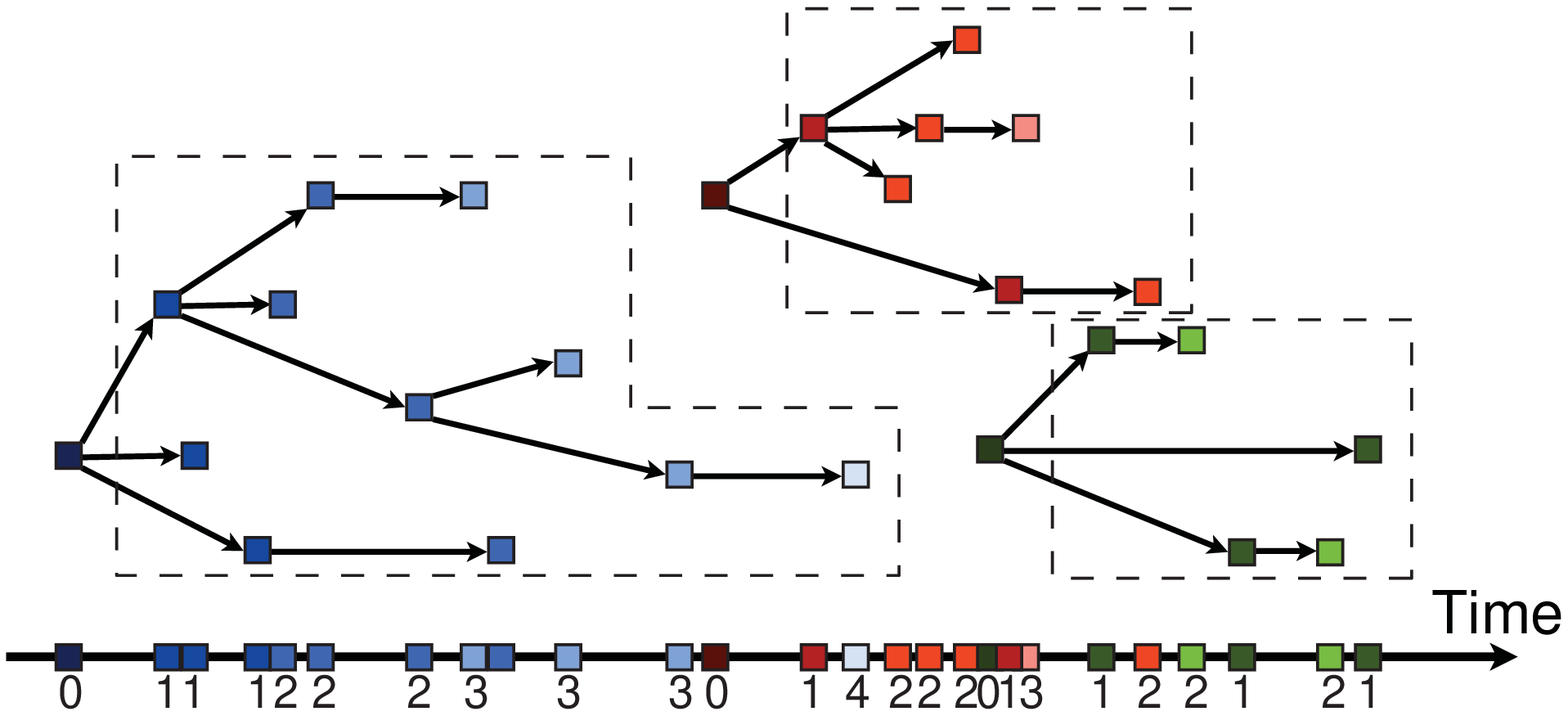}}
\caption{(Color online) Illustration of the branching structure of the Hawkes process (top) and events on the time axis (bottom). Different colors of markers correspond to different clusters, the dashed lines denote descendants of the same cluster  and the numbers next to each event denotes its order within the cluster. This picture corresponds to a branching ratio equal to $n=0.88$.}\label{fig_1}
\end{figure}

In the language of branching processes, all events belong to one of two classes --- \emph{immigrants} (or using earthquake terminology, \emph{main events}) and \emph{descendants} (or \emph{aftershocks}). The exogenous immigration (described by the background intensity $\mu(t)$) triggers clusters of descendants. Namely, every zeroth order event (immigrant) can trigger one or more first-order events, 
each of whom in turn can trigger several second-order events and so on over many generations (see fig.~\ref{fig_1}). 
All first-, second- and higher order events (descendants) form the cluster of aftershocks of the main event 
as a result of the self-excited (endogenous) generating mechanism of the system. 

The crucial parameter of the branching process is the so-called \emph{branching ratio} ($n$), which is defined
as the average number of daughter events per mother event. There are three regimes: (i) \emph{sub-critical} ($n<1$), 
(ii) \emph{critical} ($n=1$) and (iii) \emph{super-critical} or explosive ($n>1$). Starting from a single 
mother event (or immigrant) \correction{at time $t_1$}, the process dies out with probability $1$ in the sub-critical and critical regimes
and has a finite probability to explode to an infinite number of events in the super-critical regime. 
The critical regime for $n=1$ separates the two main regimes and is characterized by power law statistics
of the number of events and in the number of generations before extinction \cite{Saishorhelman}. For $n\leq 1$, the process
is stationary in the presence of a Poissonian or more generally stationary flux of immigrants.

In the sub-critical regime, in the case of a constant background intensity ($\mu(t)=\mu=\mbox{const}$),
the branching ratio is equal to the proportion of the average number of
descendants in the whole population~\cite{Sornette2003geo}. 
In other words, the branching ratio is equal to the proportion of the average
number of endogenously generated events among all events.
To see this, since by definition $n$ is the average number of first-generation events per immigrant, the total average number of 
events of all generations triggered by a given mother event is $n + n^2 + n^3 + ....= n/(1-n)$. 
The fraction of such triggered events to the total number of events (that therefore includes the mother event) is thus
the ratio of $n/(1-n)$ to $1+n/(1-n)$, which is equal to $n$. Using
the branching property that the generations of events associated to different mother events are 
forming independent ``branches'', the ratio $n$ holds for each cluster so that
the average ratio of all triggered events by all immigrants to the total number of events is equal to $n$.
Calibrating $n$ on the data therefore provides a direct quantitative estimate of the 
degree of endogeneity. 

There are several routes to determine $n$. One is to reverse-engineer the clusters and calculate
the average number of direct descendants to any given event. This can be done via the stochastic declustering
(parametric \cite{Zhuangetal02} and non-parametric \cite{Marsan07}) method, which amounts to reconstruct from the sequence of events (fig.~\ref{fig_1} bottom) the original cluster structure (fig.~\ref{fig_1} top), or at least distinguish between descendants and immigrants, but this 
may have severe limitations \cite{Sorutkin}.
A simpler way is to just use the definition
\begin{equation}\label{n}
	n=\int_0^\infty h(t)dt~.
\end{equation}
Given the parametric form of the kernel $h(t;\hat\theta)$, one can estimate the parameters 
$\hat\theta$ for the given realization of the process $t_1,\dots,t_N$, using the Maximum Likelihood Estimation
method, which benefits from the fact that the log-likelihood function
is known in closed form for Hawkes processes \cite{Ogata1978,Ozaki1979}. 

\section{Calibration to mid-price changes data and goodness of fit analysis}

For the present application,  we use the classical model \eqref{hawkes} 
and assume (i) a constant background intensity $\mu(t)=\mbox{const}$ with (ii) an exponential kernel~\cite{Bowsher2002,Bowsher2007,Toke2011,Cont2011,Muzy2010hawkes,Muzy2011}: $h(t)=\alpha e^{-\beta t}\xi(t)$, where $\beta>0$ and $\xi(t)$ is the Heaviside function that ensures causality. Given this parametrization, the branching ratio~\eqref{n} is given by $n=\alpha/\beta$. Choosing $n$ as an independent parameter and substituting it into expression \eqref{hawkes}, we finally write the conditional intensity
\begin{equation}\label{hawkes_exp}
	\lambda_t(t)=\mu+n\beta\sum_{t_i<t}e^{-\beta(t-t_i)},
\end{equation}
where $t_i$ are timestamps of the individual events and $\mu,n,\beta$ are parameters of the model.
In this formulation, all events (changes of mid-price) have identical impact on the conditional intensity, independently of the direction and size of mid-price changes or volumes at the bid or ask sides. This simplification can be removed by considering the marked Hawkes process, with marks being functions of the volumes or increments of prices. However, residual analysis indicates that 
the selected model~\eqref{hawkes_exp} fits the data with an excellent precision, and the extension to marked
or multivariate Hawkes processes is not necessary.

The assumption that the background intensity is constant implies for $n \leq 1$ that the 
events time series is stationary. As pointed earlier, the activity (the flow of events) is non stationary and, in particular, subjected to intraday seasonality. \correction{To address this issue, one needs to consider the smallest possible intervals. But, decreasing the 
size of the intervals decreases the number of events that are used for the estimation and thus its robustness. More important, the size of 
the time window limits the memory of the endogenous process taken into account in the estimation procedure.
In other words, considering time intervals of just a few minutes prevents capturing the memory effects that are developing over the scale of hours. In the present work, we resolve this trade-off by considering time intervals of a few tens of minutes (namely, windows of 10, 20 and 30 minute). In such short time intervals, the parameters of the Hawkes model~\eqref{hawkes_exp} can be considered approximately constant. At the same time, these intervals are wide enough to capture the endogenous memory of the system due to the algorithmic and high-frequency trading that operate at the scales of seconds to milliseconds.}
Moreover, windows of 10 minutes or more contain typically more
than 100 events, which allows a reliable calibration (there were on average 150 events
per 10 minutes in 1998,  350 in 2010 and 890 in 2008).  

Another characteristic of the data needs to be addressed before calibration. 
Due to timestamps being rounded to the nearest second, the dataset contains multiple events with equal timestamps. 
On average, there are approximately 0.26 events (mid-price changes) per second in 1998, and
1.5 events per second in 2008. However, during the so-called ``Flash-crash'' event (May 6, 2010 14:45 EST),
there was a peak of  194 mid-price changes per second. The Flash-crash event
occurred as algorithmic and high-frequency traders in S\&P 500 E-mini futures contracts triggered 
a dramatic fall in other markets~\cite{FlashCrash2010_report} (see fig.~\ref{fig_3}). 
To address this issue of multiple events per timestamp which may bias the calibration of the model, 
we round the timestamps by randomly redistributing events with the same timestamps
within each second interval. This amounts to assuming that each event occurring within one second
is independent of all the others within the same second interval (but not
between different seconds). To verify that this procedure does not bias the calibration of the Hawkes model, we have 
tested it on synthetic time series obtained by numerical simulation of the Hawkes process~\eqref{hawkes_exp} 
with parameters $(\mu,n,\beta)$ close to the calibrated values on the real data.
Our simulations show that the estimation errors, which result from
the finite sample size and the rounding of timestamps, are low
(standard deviations of $1.4\%$ for $\mu$, $0.6\%$ for $n$ and $3.5\%$ for $\beta$). 
Additionally, after the estimation of the parameters of the real time series, 
we performed 50 bootstraps obtained by 50 different realizations of the randomization
within each second in each time-window in 1998--2010, allowing us to 
study error bars of the estimations of $(\mu,n,\beta)$. The standard deviation of the estimated parameters is relatively small and decreases in time with increase of the number of events per time window. For instance, 
the standard deviation of the estimation of $n$ was approximately 0.13--0.18 (in absolute value) in 1998, 0.05--0.07 in 2000 and after 2002 it never exceeded the level of 0.035, decreasing by 2010 to the range of 0.005-0.015, implying a 1\%--2\% estimation error.

\begin{figure}[p]
\centerline{\includegraphics[width=0.7\textwidth]{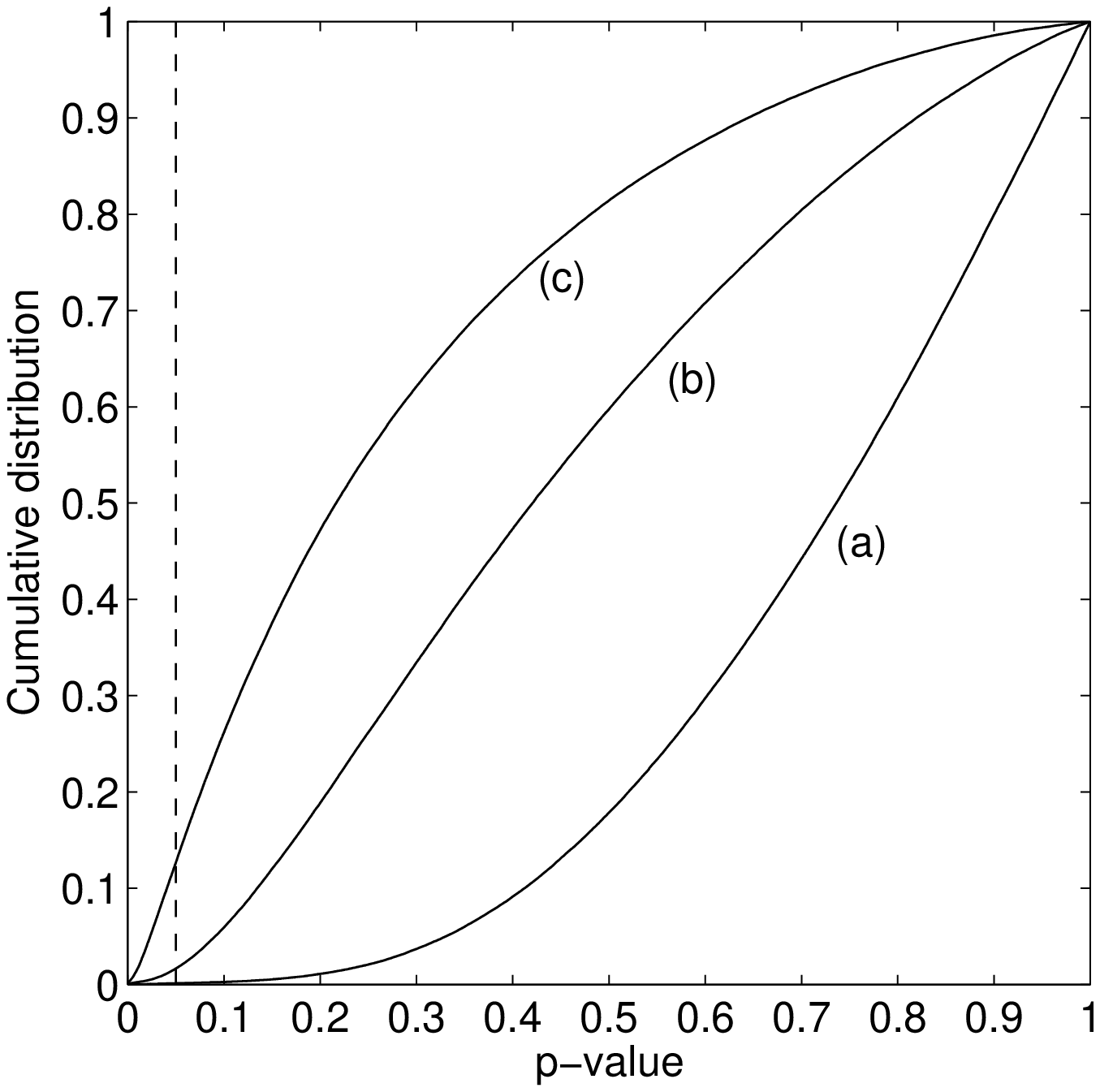}}
\caption{\correction{Cumulative distribution of the p-values obtained in the test for uniform distribution (see text) used to estimate the Hawkes model, performed in windows of (a) 10 minutes, (b) 20 minutes and (c) 30 minutes. The vertical dashed line denotes the $5\%$ threshold at which the 
statistical test performed on the calibration of the model in a given window leads to its rejection.}}\label{fig_5}
\end{figure}

The standard quantification of the goodness-of-fit of the data by the Hawkes process uses
residual analysis~\cite{Ogata1988}, which consists in studying the so-called residual process 
defined as the nonparametric transformation of the initial time-series $t_i$ 
into $\xi_i=\int_0^{t_i}\hat\lambda_t(t)dt$, where $\hat\lambda_t(t)$ is the conditional intensity of the Hawkes process
(\ref{hawkes_exp}) estimated with the maximum likelihood method. Under the null hypothesis
that the data has been generated by the Hawkes process~\eqref{hawkes_exp}, 
the residual process $\xi_i$ should be Poisson with unit intensity~\cite{Papangelou1972}. 
Visual cusum plot and Q-Q plot analysis show the excellent explanatory power of the Hawkes model with respect to the data. \correction{Visual analysis was complemented with rigorous statistical tests. Under the null hypothesis of agreement of data with the model (Poisson statistics of the residual process $\xi_i$), the inter-event times in the residual process $\Delta_i=\xi_i-\xi_{i-1}$ should be exponentially distributed with CDF $F(\Delta)=1-\exp(-\Delta)$. Thus, the random variables $U_i\equiv F(\Delta_i)=1-\exp(-\Delta_i)$ should be uniformly distributed in $[0,1]$. We have performed rigorous Kolmogorov-Smirnov tests for uniformity} for each 
calibration of the Hawkes process over 1998--2010 in moving time windows of 10 minutes 
(198'713 estimates), 20 minutes (193'877 estimates) and 30 minutes (188'909 estimates) that are swept through the Regular Trading Hours with a step of 5 minutes. \correction{We rejected the model in a given time interval if the hypothesis of a uniform distribution of the residuals $U_i$ could be rejected at 
the $5\%$ confidence level for each of the 50 different estimations obtained with bootstrapping performed for that time interval. With this criterion, we have found that, out of all 10 minutes intervals, 240 intervals ($0.12\%$) could not be described with the Hawkes model~\eqref{hawkes_exp} with exponential kernel. For 20 minutes intervals, we could reject 3'180 calibrations ($1.64\%$) and, for 30 minutes intervals, 23'878 calibrations ($12.64\%$) could be rejected. Fig.~\ref{fig_5} presents the empirical cumulative distribution of the maximal p-values for estimations performed in windows of different sizes. The larger number of rejected calibrations obtained for longer time windows results from fits that are generally poorer due to intraday non-stationarity of the data. Extending the time window size to one hour, we find that more than $50\%$ of all calibrations are rejected at the $5\%$ confidence level. Working with these short time intervals of 10 to 30 minutes, we exclude from our following analysis the small fractions of windows for which the data could not be described with the model~\eqref{hawkes_exp} at the showing threshold of 5\%. 
Overall, different tests presented in this section} confirm that (i) the chosen exponential kernel~\eqref{hawkes_exp} describes the data well, (ii) the random redistribution of the timestamps within each second does not affect the results of the estimation procedure and (iii) the assumption that the parameters are constant in each window is valid.

\section{Branching ratio $n$ and level of endogeneity in US financial markets}

\begin{figure}[p]
\centerline{\includegraphics[width=0.8\textwidth]{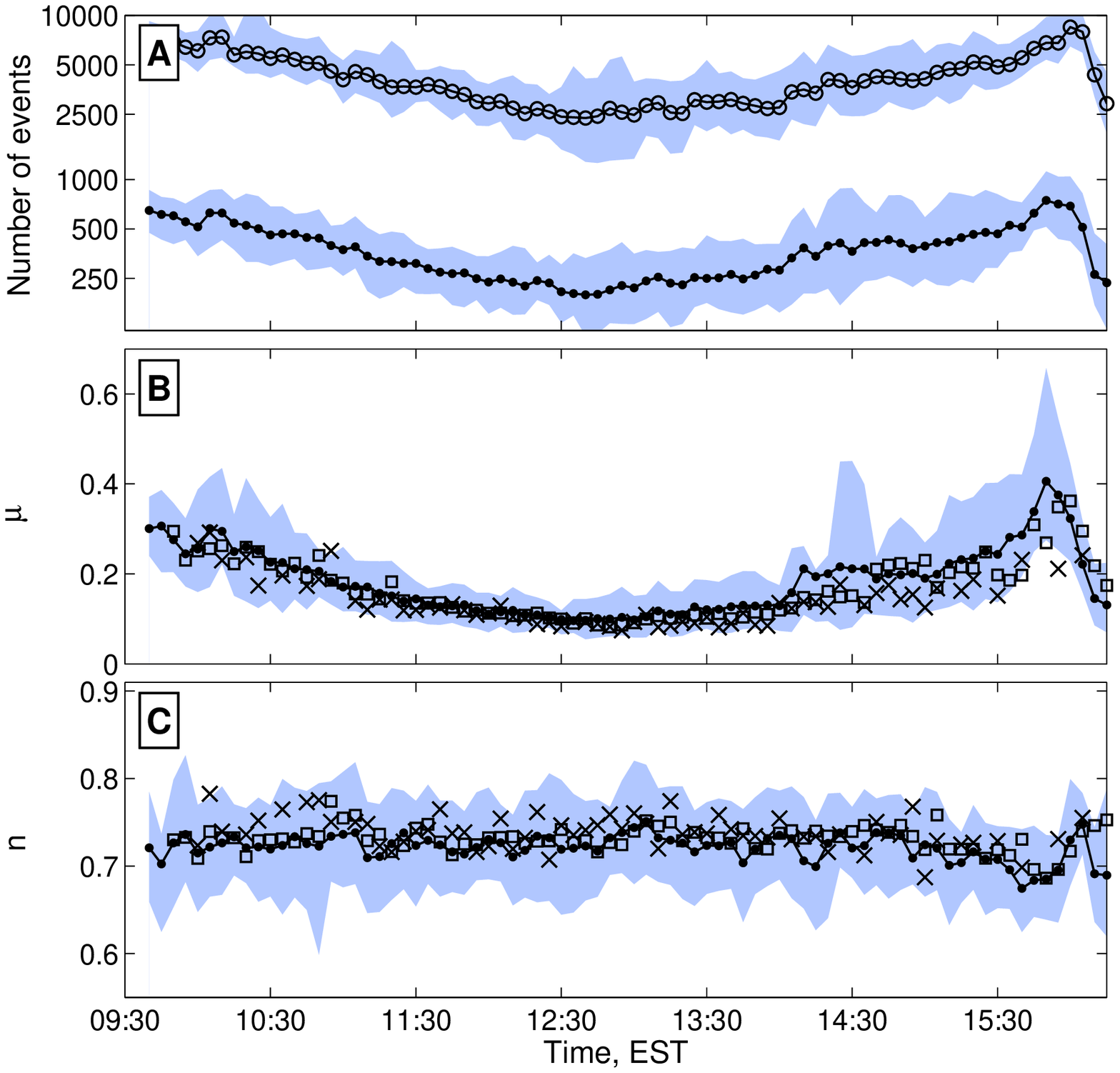}}
\caption{\correction{(Color online) (A) Number of mid-price changes (dots) and transactions (circles) for the E-mini S\&P 500 futures contracts in time windows of 10 minutes during the Regular Trading Hours in March 2009 (logarithmic scale on vertical axis). Each point represents the averaged value over the period from March 1, 2009 to March 31, 2009, and the shaded area correspond to the 10\%--90\% quantile range obtained with the same one month of data. (B) Estimated background intensity ($\hat\mu$) and (C) branching ratio ($\hat n$) of the flow of mid-price changes of the E-mini S\&P 500 futures contracts over the Regular Trading Hours. Each point at a given time $t$ represents an estimate in windows of 10 (dots), 20 (squares) and 30 (crosses) minutes averaged over the period of 1 month. The shaded area corresponds to 10\%--90\% quantile range obtained with the same 1 month of estimates for 10 minutes time windows.}}\label{fig_4}
\end{figure}

To analyze the \correction{short-term reflexivity of the market,} 
we use the maximum likelihood estimator~\cite{Ogata1978,Ozaki1979} to calibrate the Hawkes model~\eqref{hawkes_exp} 
in time windows of 10, 20 and 30 minutes spanning every day from 1998 to 2010 \correction{with a 5 minutes time step}. We excluded
the days when trading was closed before 16:15 EST or with nonactive trading. 
We also filtered out the trading days with daily volume less than the 5\% quantile of daily volumes for each given year. \correction{Fig.~\ref{fig_4} illustrate an example of the intraday behavior of the parameters in March 2009. Panel~\ref{fig_4}A illustrates the U-shape of the intraday seasonality in the trading activity discussed above: one can observe that the number of transactions and the number of mid-price changes (as well as trading volume and the numbers of market and limit orders not presented in the figure) drop by almost 50\% over the lunch time in comparison with the opening and closing levels. This seasonality is observed in the estimated background intensity $\mu$ of exogenous events in the market (panel~\ref{fig_4}B). In contrast to the background intensity, the branching ratio (panel~\ref{fig_4}C) that captures the endogenous impact of reflexivity does not exhibit a U-shape. As one can see from the figure, the branching ratio fluctuates around some mean value, showing sometimes non-regular excursions that will be 
investigated further below in two case-studies discussed in section~\ref{forecast}. While the average intraday pattern of the branching ratio can be considered as approximately constant in a first approximation, its shape slightly varies from year to year. We attribute this effect to the evolution of the trading algorithms that are the main source of the short-term reflexivity measured by $n$. Finally, we stress the good coincidence of results 
obtained from the estimations performed in windows of different sizes, which supports that the proposed method is robust.}

\begin{figure}[p]
\centerline{\includegraphics[width=0.7\textwidth]{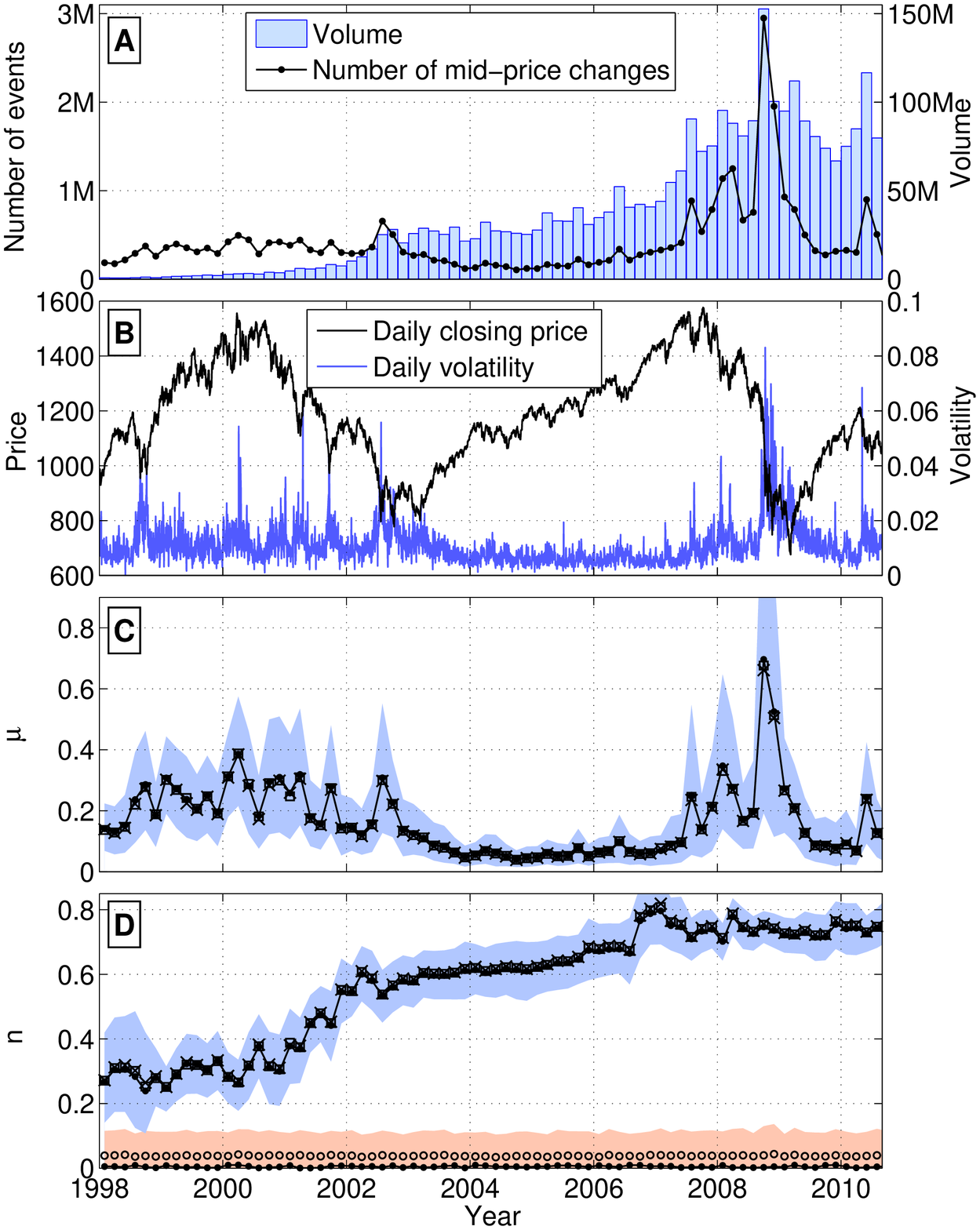}}
\caption{(Color online) (A) Number of mid-price changes (in millions of events) and volume (in millions of contracts) for the E-mini S\&P 500 futures contracts over the period 1998--2010. Each point at a given time $t$ represents the total volume and number of events over a two month interval
centered on that time $t$.  (B) Daily closing price and daily volatility \correction{estimated with the Garman\&Klass estimator~\cite{GarmanKlass1980}}. (C) Estimated background intensity ($\hat\mu$) and (D) branching ratio ($\hat n$) of the flow of mid-price changes of the E-mini S\&P 500 futures contracts over the period 1998--2010. Each point at a given time $t$ represents an averaged over a two month interval
centered on that time $t$ of windows of 10 (dots), 20 (squares) and 30 (crosses) minutes (the corresponding lines are almost indistinguishable). The shaded area corresponds to 10\%--90\% quantile range obtained with the same 2 months of daily estimates for 10 minutes time windows. \correction{The horizontal shaded area in the inset (D) corresponds to the 90\% quantile of estimations performed with the ``reshuffled'' time series (see text). Dots correspond to the median value and circles to arithmetic mean}.}\label{fig_2}
\end{figure}

\correction{To analyze the evolution of the parameters over the whole period of 1998--2010, we have} averaged the estimates for the parameters $(\mu,n,\beta)$ over all windows within a two month period.
The curves shown in fig.~\ref{fig_2} give the average values of the parameters as a function of the middle time $t$
over the time intervals $[t-\Delta_m, t+\Delta_m]$, where $\Delta_m=1$ month. 
We also determined the quantiles over the set of all 10, 20 and 30 minutes windows in 
each two month interval. Fig.~\ref{fig_2}C and D show that the results obtained for the 10, 20 and 30 minutes windows
are practically undistinguishable. This observation together with the narrowness of the
10\%-90\% quantile range, especially for the estimation of the branching ratio, 
confirm both the status of the Hawkes model as excellent description of the data
and the robustness of our estimation procedure.

Comparison of Fig.~\ref{fig_2}A-C shows that (a) the number of mid-price changes (panel A), 
(b) the daily volatility (panel B)
and (c) the background intensity $\mu$ (panel C) are behaving similarly with coincident major peaks associated with major phases
of market instabilities, during and following the burst of the ICT dotcom bubble \cite{JohSor2000} and 
associated with the financial crisis that started in 2007
and culminated with Lehman Brothers bankruptcy \cite{Sorwood2010}. 
Note that the increase of trading activity from 1998 to 2010, as proxied by volume (Fig.~\ref{fig_2}A)
is not accompanied by an increase of the background intensity $\mu$ of exogenous \correction{events} in the market.
This makes intuitive sense since $\mu$ should reflect the genuine news impacting the market. 

In contrast, the time evolution of the branching ratio $n$ presented in fig.~\ref{fig_2}D exhibits a very different behavior.
The first important observation is that, since 2002, $n$ has been consistently above 0.6 and, since 2007, between $0.7$ and $0.8$
with spikes at $0.9$. These values translate directly into the conclusion that, since 2007, more than $70\%$ of the 
price moves are endogenous in nature, i.e., are not due to exogenous news but result from positive
feedbacks from past price moves. 

\correction{It should be emphasized that the increase of the branching ratio over the period 1998--2010  is not due to 
an increase of the trading activity (measured in number of transactions or volume) over the same period. 
Neither transactions, nor volume enter directly into the formulation of the model, since individual transactions do not necessary result in a change of the mid-price. As a simple example, doubling the number of transactions by splitting each of them into two independent transactions (to keep the daily volume constant) does not affect the dynamics of the mid-price at all. Similarly, keeping the number of transactions constant and doubling the volume of each of them (doubling the volume of each incoming market order) while simultaneously doubling the volume of all incoming limit orders again would not change the dynamics of the mid-price. The ``decoupling'' of market transactions from the simple measure of activity can be seen from fig.~\ref{fig_2}A, where the dramatic increase of volume in 1998--2007 was not accompanied by an increase of the number of mid-price changes. In fact, the latter decreased in 2003--2005.} 

\correction{In order to reject the possibility that the observed 
increase of the branching ratio over 1998--2010 is due to the increase of the trading activity over that period,
we have performed the following test. We have fixed the number of mid-price changes per day, but have redistributed these events such that, within one day, their dynamics was described by a Poisson process. This ``reshuffling'' of the time series amounts to keeping the price trajectories, the daily volume, the number of price and mid-price changes per day unchanged (i.e., keeping the same trajectories as shown in fig.~\ref{fig_2}A and B), while
distorting time such that the intervals between consecutive mid-price changes within one day become uncorrelated and exponentially distributed. 
We have then performed exactly the same procedure as described above (dividing each day in 10 minutes interval and estimating the parameters of the Hawkes process~\eqref{hawkes_exp}) within these intervals. As one can see from the bottom shaded area in fig.~\ref{fig_2}D that represents
the 90\% quantile of the branching ratio estimated within such distorted time series, one can clearly reject the hypothesis 
that the branching ratio is sensitive to, or equivalently provides another measure, of trading activity. This quantitative result supports the key property of the Hawkes model, which is that the branching ratio is not determined by the average rate of events but by the degree of self-excitation of the system.}

\correction{In addition to the increase of the level of endogeneity in 1998--2010, one can observe another remarkable fact in Fig.~\ref{fig_2}D 
corresponding to} the existence of four market regimes over the period 1998-2010:
\begin{itemize}
\item[(i)] In the period from Q1-1998 to Q2-2000, the final run-up of the dotcom bubble is associated
with a stationary branching ratio $n$ fluctuation around $0.3$. 
\item[(ii)] From Q3-2000 to Q3-2002, $n$ increases from $0.3$ to $0.6$. This regime corresponds
to the succession of rallies and panics that characterized the aftermath of the
burst of the dot-com bubble and
an economic recession \cite{SorZhoudotcomafter2002,Zhousorren2003}.
\item[(iii)] From Q4-2002 to Q4-2006, one can observe a slow increase of $n$ from $0.6$ to $0.7$.
This period corresponds to the ``glorious years'' of the twin real-estate bubble,
financial product CDO and CDS bubbles, stock market bubble and commodity bubbles  \cite{Sorwood2010}.
\item[(iv)] After Q1-2007 the branching ratio stabilized between $0.7$ and $0.8$ corresponding to the start of the problems of the subprime financial crisis (first alert in Feb. 2007), whose aftershocks are still resonating at the time of writing.
\end{itemize}

As already mentioned, the value of the branching ratio larger than 0.7 since 2007 indicates 
that more than 70\% of the price movements can be attributed to endogenous processes occurring in the market. 
Notwithstanding such rather large calibrated branching ratios $n$, they are most likely underestimations
of the real values.  Indeed, 
the Hawkes self-excited model has been calibrated on short intra-day time windows, using
a short-memory exponential kernel $h(t)$. This means
that price moves before any given time window of 10, 20 or 30 minutes that could trigger
price changes within the window are not taken into account. This truncation is known to decrease
artificially the observed branching ratio $n$ and increase the background rate $\mu$ \cite{SorWerner2005}.
In other words, neglecting past events before the short-term windows and their triggering effect leads to the
misattribution that many of the endogenous events are exogenous.

The endogeneity of $n \geq 70\%$ that we observe in 
the short time windows (10, 20 and 30 minutes) 
captures short-term feedback mechanisms within financial systems that 
can be interpreted as short-term reflexivity of humans and algorithmic trading systems.
Indeed, the  growth of the branching ratio $n$ coincides with the appearance and dramatic growth of algorithmic and high-frequency trading, 
whose birth is usually dated to 1998 when the U.S. Securities and Exchange Commission (SEC) authorized electronic exchanges. 
In the early 2000s, high-frequency trading (defined as the high speed component of algorithmic trading) was quite rare 
and accounted for less than 10\% of all equity orders. In subsequent years, its importance grew rapidly, increasing by about 164\% between 2005 and 2009~{\cite{Duhigg2009_TNYT}}. In 2009, the proportion of high frequency trading in US markets was estimated 
as more than 60\% by the TABB Group~{\cite{TABB_HFT_2009}} and the Aite Group~\cite{Aite_HFT_2009}. 
Thus, there is no contradiction between a rather low value $n \approx 0.3$ during 
the final run-up of the dotcom bubble from Q1-1998 to Q2-2000
and the strong herding that is often invoked to explain its development.
As said above, our calibration refers to short-term endogeneity at the time scale of 10 minutes, while
the herding mechanism thought to be at the origin of the dotcom bubble has been operating at time
scales of years \cite{Sornettecrashbook,KaizokiSorreview}.

\section{Diagnostic of criticality: towards the forecast of flash crashes}\label{forecast}

We now show with the famous example of the ``flash-crash'' of May 6, 2010 that the calibration of the level
of endogeneity $n$ may provide instantaneous characteristic signatures of anomalous market regimes, 
which could be used as precursors for forecasts.
  
  \begin{figure}[p]
\centerline{\includegraphics[width=0.8\textwidth]{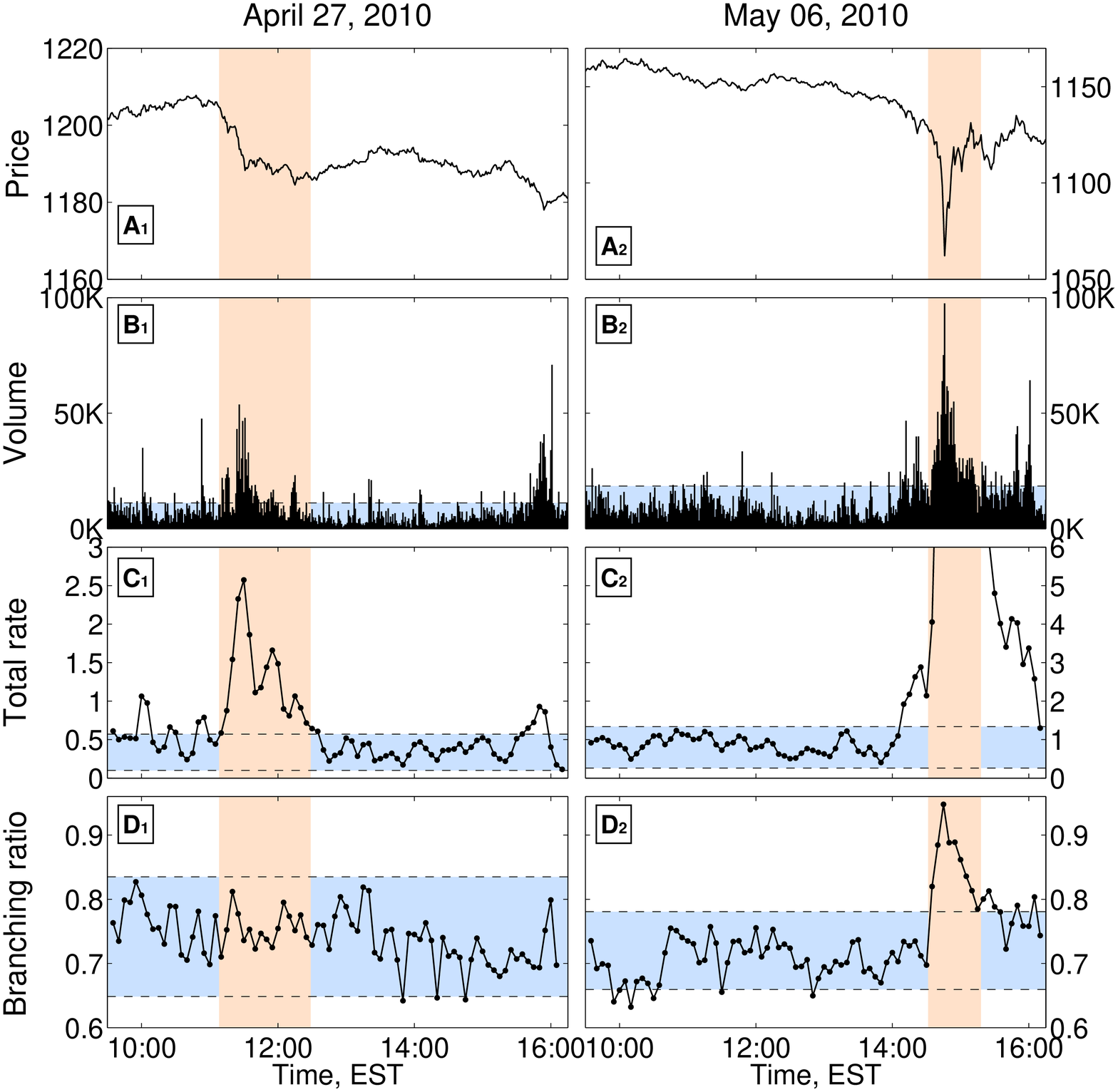}}
\caption{(Color online) Case study of two extreme events that occurred during the spring of 2010. The left column corresponds to April 27, when almost all US markets fell significantly following the dramatic decrease of the credit rating of Greece and Portugal~\cite{CNNMoney_27Apr2010}. The right column  corresponds to May 6, 2010, the so-called ``flash-crash'' event, when the activity of high-frequency traders of the S\&P 500 E-mini futures contracts leaded to a dramatic fall in other markets~\cite{FlashCrash2010_report}. Panels from top to bottom: Minute-by-minute price (A) and volume (B) of S\&P 500 E-mini futures contracts (maturity: June, 2010). (C) Average number of mid-price changes per second (averaged in 10 minutes interval). (D) Branching ratio $n$ estimated within each 10-minutes interval. The blue horizontal regions depict the 5\%--95\% quantile intervals of the corresponding values calculated over the previous trading day. The pink vertical regions highlight the time periods of the most active trading. }\label{fig_3}
\end{figure}

For this, we consider two extreme events that occurred in the spring of 2010 that are comparable
in their price amplitudes and market-wide impacts, as can be seen from the 
top four panels of Fig.~\ref{fig_3}. On April 27, 2010, 
all major US indices and stocks felt significantly after Standard\&Poors cut Greece's debt rating to ``BB+'' and lowered Portugal's debt rating, raising fears that a euro zone debt crisis could slow the global economic recovery~\cite{CNNMoney_27Apr2010}. 
On May 6, \correction{2010}, in a general atmosphere of pessimism and worry concerning the debt crisis in Greece, a large
market order triggered a flurry of activities by algorithmic traders resulting in 
a large drop of the price of S\&P 500 E-mini futures. Due to the coupling with other markets
via hedging and portfolio effects, this drop cascaded to many other markets, 
triggering drawdowns of up to 60\% in some of them~\cite{FlashCrash2010_report}. 

The top four panels of Fig.~\ref{fig_3} show that the two extreme events of April 27 and May 6, 2010
have similar price drops and volume of transactions. In particular, we find that the 
volume  was multiplied by 4.7 for April 27, 2010 and by 5.3 times for May 6, 2010
in comparison with the 95\% quantile of the previous days' volume. The main difference lies in the trading
rates and in the branching ratio. Indeed, the event of April 27, 2010 can be classified according to our
calibration of the Hawkes model as a pure exogenous event, since the branching ratio $n$ (fig.~\ref{fig_3}D$_1$) does not exhibit any statistically significant change
compared with previous and later periods. In contrast, for the May 6, 2010 flash crash, one can observe a statistically significant 
increase of the level of endogeneity $n$ (fig.~\ref{fig_3}D$_2$).
At the peak, $n$ reaches $95\%$ from a previous average level of $72\%$, which means that, at the peak (14:45 EST), 
more than $95\%$ of the trading was due to endogenous triggering effects rather than genuine news. 

Comparing the trading rates in panels C$_1$ and C$_2$ of Fig.~\ref{fig_3}, one can observe that until time 14:30 EST the trading rate of May 6 was increasing at a similar rate as for April 27.  
Therefore, on the basis of trading rates, it would not have been possible to predict the subsequent
jump in trading rate that occurred during the flash crash. But the panel D$_2$ for $n$
shows that, at this time, the instantaneously statistical estimation of the branching ratio $n$
was already giving an abnormal reading, in the sense that $n$ jumped above the $95\%$ quantile.
In contrast, for the April 27, 2010 event, the branching ratio was fluctuating normally within 
its normal band. 

This comparison suggests that the estimation of the branching ratio
provides a novel powerful metric of endogeneity, which is much richer than standard
direct measures of activity such as volume and trading rates. Indeed, the branching ratio
provides a direct access to the level of endogeneity of the market. The distance of $n$ to the critical
value $1$ can be taken as a gauge of the degree to which the market is going ``critical''.
The term actually is more than just suggestive: ``criticality'' in nuclear reactions precisely
refers to a branching process of neutrons triggering and being created by nuclear reactions
for which the process does not stop but may in fact explode. Similarly, as $n$ approaches $1$,
we can state that the market approaches ``criticality'' in this precise sense of a theoretically diverging
trading activity in absence of any external driving. 

\section{Conclusion}

\correction{We have provided what is, to the best of our knowledge, the first quantitative
estimate of the degree of reflexivity, measured as the proportion $n$ of price moves due to 
endogenous interactions to the total number of all price moves that also include the impact of exogenous news.
For this, we have used the self-excited conditional Poisson
Hawkes model \cite{Hawkes1971}, which combines in a natural and parsimonious way exogenous influences with self-excited dynamics.
Within the Hawkes model framework, the parameter $n$ takes the simple meaning of being the 
average branching ratio or, equivalently, the average number of triggered events of first generation per exogenous source.
We have calibrated the Hawkes model to the E-mini S\&P 500 futures contracts traded in the Chicago Mercantile Exchange from 1998 to 2010. 
We find that the level of endogeneity has increased significantly from 1998 to 2010, with only 70\% in 1998 to less than 30\% since 2007 of the price changes resulting from some revealed exogenous information. We have also documented
a drastic difference in the change of $n$ before and during two extraordinary flash crash events that occurred respectively 
on April 27, 2010 and on May 6, 2010. For the former, we find that the branching ratio $n$ remained constant, exemplifying
the exogenous nature of the crash. For the later, in contrast, we document both a precursory early rise of $n$
followed by strong increase culminating very close to the critical value $n=1$, suggesting a strong endogenous component.
}

In conclusion, the present study enlarges considerably the usefulness and operational implementation 
of the Hawkes model, which was already used in the study of the dynamics of book sales
\cite{Sornettedeschatres}  and of views of YouTube videos  \cite{Sornettecrane}.
Indeed, these two previous studies have selected the blockbusters, which are characterized
by large social and/or attention. This amounts to introducing a selection bias towards dynamics (i.e. those books
and videos) that are close to critical ($n \simeq 1$), as confirmed by the  pure power law behavior~\cite{footnote_1} documented in \cite{Sornettedeschatres,Sornettecrane}. In contrast, the present study
has demonstrated how to quantify the level of endogeneity $n$, which is found to characterize
different social regimes. Our work opens the road towards the full utilization of the 
dynamics of $n$ in order to diagnose different regimes and to possibly forecast 
impending crises associated with the approach to criticality $n \geq 1$.

\noindent {\bf Acknowledgments}:
We would like to thank Georges Harras for valuable suggestions and discussions during 
the analysis of the data and preparation of the paper. We are also grateful to Prof. Alexander Saichev and Dr. Mika Kastenholz for
fruitful discussions.

\pagebreak

\end{document}